\documentclass[aps,prd,reprint,groupedaddress,preprintnumbers,longbibliography]{revtex4-1}

\usepackage{amsmath,amssymb,bm,mathrsfs}
\usepackage{graphicx,color}
\usepackage{amsfonts}
\usepackage[export]{adjustbox}
\usepackage{mathtools}
\usepackage[sort&compress]{natbib}
\usepackage{srcltx}
\usepackage{dsfont}
\usepackage{epsfig}
\usepackage{slashed,braket}       
\usepackage{bbold}
\usepackage[dvipsnames]{xcolor}
\PassOptionsToPackage{caption=false}{subfig}
\usepackage{subfig}
\usepackage{xfrac}
\usepackage{multirow}
\usepackage{booktabs}
\usepackage[colorlinks=true
,urlcolor=blue
,anchorcolor=blue
,citecolor=blue
,filecolor=blue
,linkcolor=red
,menucolor=blue
,linktocpage=true
,pdfproducer=medialab
,pdfa=true
]{hyperref}
\usepackage{cleveref}
\usepackage{setspace}

\newcommand{\rhoDM}{\rho_{\scriptscriptstyle \textrm{DM}}}

\newcommand{\Fig}[1]{Fig.~\ref{#1}}
\newcommand{\Refc}[1]{Ref.~\cite{#1}}
\newcommand{\Refs}[1]{Refs.~\cite{#1}}
\newcommand{\Eq}[1]{Eq.~\eqref{#1}}

\usepackage{tikz}
\usepackage{tkz-euclide}
\usetikzlibrary{decorations.pathmorphing}	
\tikzset{
    v/.style={decorate, decoration={snake, segment length=3mm, amplitude=0.75mm}, draw},
    f/.style={draw,decoration={markings,mark=at position #1 with {\arrow[very thick]{latex}}},postaction={decorate},node contents=#1},
    f/.default=.6,
    fb/.style={draw,decoration={markings,mark=at position #1 with {\arrowreversed[very thick]{latex}}},postaction={decorate},node contents=#1},
    fb/.default=.4,
    fnar/.style={draw},
    g/.style={decorate, draw,  decoration={coil,amplitude=3pt, segment length=3.5pt}},
    s/.style={dashed,draw, postaction={decorate},
        decoration={markings,mark=at position .55 with {\arrow[very thick]{latex}}}},
    sb/.style={dashed,draw, postaction={decorate},
        decoration={markings,mark=at position .55 with {\arrowreversed[draw=black,very thick]{latex}}}},
    snar/.style={dashed,draw,line width =1.25pt},
}
\usetikzlibrary{shapes}	
\tikzset{every picture/.style={line width=1}}

\definecolor{c1}{RGB}{184, 13, 72}\colorlet{c1}{c1!80!black}
\colorlet{c1A}{c1!70!white}
\colorlet{c1B}{c1!85!white}
\colorlet{c1C}{c1!85!black}
\colorlet{c1D}{c1!70!black}

\definecolor{c2}{RGB}{43, 106, 108}
\colorlet{c2A}{c2!70!white}
\colorlet{c2B}{c2!85!white}
\colorlet{c2C}{c2!85!black}
\colorlet{c2D}{c2!70!black}

\definecolor{c3}{RGB}{242, 151, 36}
\colorlet{c3A}{c3!50!white}
\colorlet{c3B}{c3!75!white}
\colorlet{c3C}{c3!75!black}
\colorlet{c3D}{c3!50!black}

\begin{document}

\preprint{CERN-TH-2022-163}
\preprint{DESY 22-155}

\title{On the Sensitivity of Spin-Precession Axion Experiments}

\author{Jeff A. Dror}
\affiliation{Department of Physics, University of California Santa Cruz, 1156 High St., Santa Cruz, CA 95064, USA\\
and Santa Cruz Institute for Particle Physics, 1156 High St., Santa Cruz, CA 95064, USA}

\author{Stefania Gori}
\affiliation{Department of Physics, University of California Santa Cruz, 1156 High St., Santa Cruz, CA 95064, USA\\
and Santa Cruz Institute for Particle Physics, 1156 High St., Santa Cruz, CA 95064, USA}

\author{Jacob M. Leedom}
\affiliation{Deutsches Elektronen-Synchrotron DESY, Notkestr. 85, 22607 Hamburg, Germany}

\author{Nicholas L. Rodd}
\affiliation{Theoretical Physics Department, CERN, 1 Esplanade des Particules, CH-1211 Geneva 23, Switzerland}

\begin{abstract}
\noindent We study the signal and background that arise in nuclear magnetic resonance searches for axion dark matter, finding key differences with the existing literature.
We find that spin-precession instruments are much more sensitive than what has been previously estimated in a sizable range of axion masses, with sensitivity improvement of up to a factor of 100 using a ${}^{129}{\rm Xe}$ sample.
This improves the detection prospects for the QCD axion, and we estimate the experimental requirements to reach this motivated target.
Our results apply to both the axion electric and magnetic dipole moment operators.
\end{abstract}

\maketitle

The axion, a light scalar whose leading interactions exhibit a shift symmetry, is one of the most compelling extensions to the Standard Model of particle physics.
Originally proposed as an elegant solution to the Strong CP problem~\cite{Peccei:1977hh,Peccei:1977ur,Weinberg:1977ma,Wilczek:1977pj}, axions have since been appreciated for both their ubiquity in string theory~\cite{Svrcek:2006yi,Arvanitaki:2009fg,Halverson:2019cmy} and the generic expectation that they contribute to cold dark matter through, for example, misalignment production~\cite{Abbott:1982af,Preskill:1982cy,Dine:1982ah}.
The continuous shift symmetry of the axion leads to a natural expectation that the axion should be an extraordinarily light state, with mass $m_a \ll 1~{\rm eV}$ suppressed by its decay constant $f_a$, which arises from breaking the continuous symmetry to a discrete one via instanton effects.
Such a small mass implies that axion dark matter on Earth should be well described by a classical wave.

At low energies, the possible interactions of an axion, $a$, with a nucleon $N$ take the following form,
\begin{equation}
{\cal L} \supset g_N  (\partial_\mu a) \bar{N} \gamma^\mu \gamma_5 N  
- \frac{i}{2} g_d a \bar{N} \sigma_{\mu\nu} \gamma_5 N F^{\mu\nu}.
\label{eq:ops}
\end{equation}
In the presence of an axion-wave background, these interactions source an oscillating magnetic dipole (MD), weighted by $g_N \propto  1/f_a$, and an oscillating electric dipole (ED), weighted by $g_d \propto 1/m_N f_a$, with $m_N$ the nucleon mass.
For the QCD axion, $g_N$ depends on the charge assignments and field content of the UV theory, whereas $g_d$ depends on the term that resolves the strong CP problem, $(a/f_a) G^{\mu \nu} \tilde{G}_{\mu \nu}$, and is fixed to $g_d = (3.7 \pm 1.5) \times 10^{-18}~{\rm GeV}^{-2} (10^{15}~{\rm GeV}/f_a)$~\cite{Pospelov:1999mv}.
For a non-relativistic nucleus, these interactions lead to the Hamiltonian, $H_{\rm int} = - 2 ( g_N  \nabla a + g_d a {\bf E}^\ast ) \cdot {\bf S}$.
Here, ${\bf S}$ is the nucleon spin operator, and ${\bf E}^\ast $ is the effective electric field felt by a nucleus (and differs from the applied field by at least two orders of magnitude due to shielding by the atomic electrons).
By drawing an analogy between the interaction of spins and a magnetic field, the axion-nucleus interaction can be characterized as an effective axion magnetic field,
\begin{equation} 
{\bf B}_a(t) = -\frac{2}{\gamma}  \left\{ \begin{array}{cr} g_N \nabla a(t) & \text{(MD)}, \\[0.1cm]
g_d {\bf E}^\ast a(t) & \text{(ED)}, \end{array} \right.
\label{eq:Ba}
\end{equation} 
with $\gamma$ being the gyromagnetic ratio of the nucleon.
This axion-induced magnetic field can be detected using spin-precession techniques such as nuclear magnetic resonance (NMR), as originally proposed in the seminal CASPEr papers~\cite{Graham:2013gfa,Budker:2013hfa}, and it was further developed experimentally and theoretically in \Refs{Stadnik:2013raa,Abel:2017rtm,Wang:2017ixp,Garcon:2019inh,Smorra:2019qfx,Roussy:2020ily,Jiang:2021dby,Aybas:2021nvn,JacksonKimball:2017elr,Aybas:2021cdk}.
The CASPEr approach is opening up a frontier for axion dark-matter direct detection beyond the widely exploited axion-photon coupling (for related proposals, see \Refs{Graham:2011qk,Sikivie:2014lha,Graham:2017ivz,Chang:2017ruk,Chang:2018dvm,Wu:2019exd,Terrano:2019clh,Bloch:2019lcy,Yang:2019xdz,Abusaif:2019gry,Flambaum:2019rvu,Stephenson:2020jzx,Gramolin:2020ict,Graham:2020kai,Gaul:2020bdq,Arvanitaki:2021wjk,Kim:2021pld,Gao:2022nuq,Berlin:2022mia,Lisanti:2021vij,Lee:2022vvb}).

In this {\em Letter}, we revisit the sensitivity of spin-precession experiments, clarifying fundamental aspects of the behavior of the expected axion signal and noise sources.
The system depends on three fundamental timescales: the axion coherence time, $\tau_a \sim (4~{\rm neV}/m _a )~{\rm sec}$, the transverse spin-relaxation time, $T_2$, and the experimental integration time, $T$.
We demonstrate that there are two previously overlooked effects that enhance the growth of the signal when $\tau_a < T_2$, a realization that improves the prospects for QCD axion detection with spin-precession instruments, and we comment on the requirements to achieve this goal.
We further reconsider a dominant noise source -- spin-projection noise -- and demonstrate that it is larger at high axion masses, thereby reducing the utility of using materials with large magnetic moments.

\vspace{0.2cm}
\noindent {\bf The NMR Axion Signal.}
%
We begin with a brief review of spin-precession axion experiments.
Fundamentally, they involve a macroscopic sample of atoms with non-zero nuclear spin placed within a static magnetic field, ${\bf B}_0$.
The field induces a bulk magnetization ${\bf M}_0$ that is parallel to  ${\bf B}_0$.
A perpendicular magnetic field -- such as that induced by the axion -- will rotate the nuclear spins by a small amount.
However, as soon as they do, the spins will precess around ${\bf B}_0$ at the Larmor frequency, $\omega_0 \equiv \gamma B_0$, generating an oscillating transverse magnetization, which can then be detected with a sensitive magnetometer.
For $m_a \sim \omega_0$, the effect is resonantly enhanced, with the width of the system's response controlled by the transverse relaxation time, $T_2$, which is a macroscopic property of the sample and quantifies how long the precession of the transverse spins can be maintained coherently.
By varying ${\bf B}_0$, the instrument can scan a range of axion masses.

To study the dynamics of the magnetization, we turn to the Bloch equations.
Preparing the sample with ${\bf M}_0 \propto {\bf B}_0 \propto \hat{\bf z}$, the Bloch equations read
\begin{equation} 
\frac{d{\bf M}}{dt} 
= {\bf M} \times \gamma {\bf B} - \frac{M_x \hat{\bf x} + M_y \hat{\bf y}}{T_2} - \frac{(M_z-M_0) \hat{\bf z}}{T_1}. 
\label{eq:Bloch}
\end{equation}
The longitudinal relaxation time commonly satisfies $T_1 \gg T_2$, and we will work in the limit where $T_1$ is much longer than any other timescale of interest, so that it will play no further role in our discussion.
We decompose the magnetic field as ${\bf B} = B_0 \hat{\bf z} + {\bf B}_a(t)$.
As $|{\bf B}_a(t)| \ll B_0$, we study the magnetization perturbatively, taking ${\bf M} \simeq {\bf M}_0 + {\bf M}_a$, and we will work only to linear order in the axion-induced fields throughout~\footnote{This is an excellent approximation.
If we take $T_2,\tau_a \to \infty$, then $M_x(t)\sim g_N v \sqrt{2\rhoDM} M_0 \sin(\omega_0 t) t$ for the MD interaction, see Eq.~(\ref{eq:Mx}).
This suggests that with enough time an infinite magnetization can be generated, which is inconsistent with the finite number of spins in the system.
Terms higher order in $g_N$ prevent this from occurring.
We can estimate when they must enter by determining when $M_x(t) \sim M_0$.
Setting $g_N$ to the largest allowed value (set by the SN 1987A bound), we find $t \sim 30~{\rm years}$, dramatically larger than $T_1$ for any sample we consider, which represents the longest time we can interrogate the sample for.
Including a finite $T_2$ or $\tau_a$ only strengthens this conclusion.
}.
Working to this order, $M_z(t) = M_0$, leaving the dynamics to the transverse magnetizations,
\begin{equation}\begin{aligned}
\dot{M}_x & = \omega_0 M_y - T_2^{-1}M_x - M_0 \gamma B_{ay}, \\ 
\dot{M}_y & = -\omega_0 M_x - T_2^{-1} M_y + M_0 \gamma B_{ax}.
\label{eq:Bloch2}
\end{aligned}\end{equation}
These equations can be decoupled.
If we measure the $\hat{\bf x}$ component of the magnetization, the equation to solve is
\begin{equation} \begin{aligned} 
&\ddot{M}_x + 2T_2^{-1} \dot{M}_x + \omega_0^2  M_x \simeq F(t), \\
\hspace{-1.5cm}\text{where,}\hspace{0.3cm}
& F(t) = \gamma M_0 [ \omega_0 B_{ax} - \dot{B}_{ay}].
\label{eq:Bloch3}
\end{aligned}\end{equation}
Here and throughout, we neglect terms of ${\cal O} (1/\omega_0 T_2)$, as they are significantly suppressed.
\Eq{eq:Bloch3} has reduced the system to the form of a simple harmonic oscillator with a resonant frequency $\omega_0$ and bandwidth $1/T_2$, that is being driven by the axion wave.
Assuming $M_x(0)=M_y(0)=0$, the solution is given by~\footnote{The solution in Eq.~(\ref{eq:Mxsol}) is for the Bloch equation before terms of $\mathcal{O}(1/\omega_0T_2)$ are dropped; cf. Eq.~(\ref{eq:Bloch3}).}
\begin{equation} 
M_x (t) = \frac{1}{\omega_0} \int_0^t dt' e^{(t'-t)/T_2} \sin \left[ \omega_0 (t-t') \right] F(t').
\label{eq:Mxsol}
\end{equation}

To complete our solution for the magnetization, we require a model for the axion field.
Here we treat the axion as a field with a fluctuating phase---we show that our results are reproduced when the axion is modeled as a sum over plane waves in the Supplementary Material (SM).
The axion is taken to have constant amplitude $a_0 = \sqrt{2\rhoDM}/m_a$, fixed by the local dark-matter density, and oscillates with frequency $\omega_a \simeq  m_a (1+v^2/2)$.
The statistics of the field are then encoded by requiring the field, which carries velocity $v \sim 10^{-3}$, obtain a new random phase uniformly sampled on $ [0,2\pi)$, every coherence time, $\tau_a = 2 \pi/m_a v^2$~\footnote{The coherence time is a measure of how long the axion field can be approximated as a perfectly coherent driving force, or in the frequency domain how long of a measurement is required to resolve the intrinsic width of the axion.
It is an approximate notion, as the transition from perfectly coherent to incoherent happens gradually near $t \sim 2 \pi/m_a v^2$, not exactly at $\tau_a$.
For further discussion of the coherence time, see for instance Ref.~\cite{Foster:2020fln}.}.
Each time the phase is updated, we further update the direction of the axion field's momentum, ${\bf k}$ (where $|{\bf k}| = m_a v$), though, parametrically, our results are insensitive to the stochastic nature of the momentum vector.
In summary,
\begin{equation}\begin{aligned}
F(t) & = A \cos[ \omega_a t + \varphi(t)], \\
A &\equiv 2 M_0 a_0 \left\{\begin{array}{l}
g_N
\sqrt{\omega_0^2 ({\bf k} \cdot \hat{\bf x})^2 + \omega_a^2 ({\bf k} \cdot \hat{\bf y})^2}, \\[0.1cm]
g_d \omega_0 E^\ast,
\end{array} \right.
\label{eq:F}
\end{aligned}\end{equation}
where we focus on the dominant term in the driving force, which carries a phase $\varphi(t) $, uniformly sampled on $\left[0 , 2\pi \right)$, but shifted from the axion phase. Furthermore, we assumed ${\bf E}^\ast \propto \hat{\bf x}$ for the ED operator.
From \Eq{eq:Bloch3}, a resonant response is induced in the system when $\left| \omega_0 - \omega_a \right| \lesssim \pi\, {\rm max}[\tau_a^{-1},T_2^{-1}]$.
We will assume a scan strategy such that this condition is always satisfied, and thereby assume $\omega_a \simeq \omega_0$.

Whenever $t \ll \tau_a$, the axion behaves as if it were a perfectly coherent driving force: $\varphi$ and ${\bf k}$ are constant, so that \Eq{eq:Mxsol} yields the following oscillating solution
\begin{equation} 
M_x(t) \simeq \left( 1 - e^{-t/T_2} \right) \frac{A T_2}{2 \omega_0} \sin [ \omega_0 t + \varphi ].
\label{eq:Mx}
\end{equation} 
For $t \ll T_2$, the amplitude grows linearly in time.
Beyond $T_2$, however, the growth saturates.

We next extend these results to finite $\tau_a$.
Due to the stochastic variation of $\varphi$ and ${\bf k}$ for $t > \tau_a$, it is useful to compute the autocorrelation function of the induced magnetization, $C(t ,t') \equiv \langle M_x(t) M_x(t')\rangle$,
\begin{equation}\begin{aligned}
C(t,t') & = \frac{1}{\omega_0^2} \int_0^t d\bar{t} \int_0^{t'} d \bar{t}' e^{-(t-\bar{t}) / T_2} e^{-(t'-\bar{t}')/T_2} \\ 
\times & \sin \left[ \omega_0 (t-\bar{t}) \right] \sin \left[ \omega_0 (t'-\bar{t}') \right] \langle F(\bar{t}) F(\bar{t}') \rangle.
\label{eq:int}
\end{aligned}\end{equation}
The expectation value vanishes unless the random phases are identical, which requires $|\bar{t}-\bar{t}'| < \tau_a$, so that
\begin{equation}
\langle F(\bar{t}) F(\bar{t}') \rangle = \frac{1}{2} \langle A^2 \rangle \cos\left[ \omega_0 (\bar{t}-\bar{t}') \right] \Theta (\tau_a - |\bar{t}-\bar{t}'|),
\label{eq:FF}
\end{equation}
in terms of the step function $\Theta$.
(Here, we have also used that $\varphi$ and ${\bf k}$ are uncorrelated.)
The remaining expectation value can be determined by averaging over the incident direction of the axion, yielding
\begin{equation}\begin{aligned}
\langle A^2 \rangle
\simeq \left( 2 M_0 a_0 \omega_0 \right)^2
\left\{ \begin{array}{l} (g_N \omega_0 v)^2/3, \\[0.1cm] (g_d E^\ast)^2. \end{array} \right.
\label{eq:Asq}
\end{aligned}\end{equation}

At short times ($t,t' \ll \tau_a$) where the axion is continuous, we have (cf. \Eq{eq:Mx}),
\begin{equation}\begin{aligned}
C(t,t') &= \frac{\langle A^2 \rangle T_2^2}{8 \omega_0^2} \cos \left[ \omega_0 (t - t') \right] \\ 
\times &\left( 1 - e^{-t/T_2} \right) \left( 1 - e ^{-t'/T_2 } \right)\!.
\label{eq:MMcont}
\end{aligned}\end{equation}
At longer times, the stochastic fluctuations in the axion associated with $\tau_a$ must be accounted for.
Combining Eqs.~\eqref{eq:int} and \eqref{eq:FF}, for $t,t' \gg \tau_a$ we can evaluate the integral by rotating coordinates to $\bar{t}\pm\bar{t}'$, yielding
\begin{equation}\begin{aligned}
C(t,t') & = \frac{\langle A^2 \rangle T_2 \tau_a}{16 \omega_0^2} \cos \left[ \omega_0 (t - t') \right] \\
\times &e^{- (t + t')/T_2} \left( e^{2{\rm min}(t, t')/T_2} - 1 \right)\!.
\label{eq:MMstoc}
\end{aligned}\end{equation}

\begin{figure*}
\begin{center} 
\includegraphics[width=0.47\textwidth]{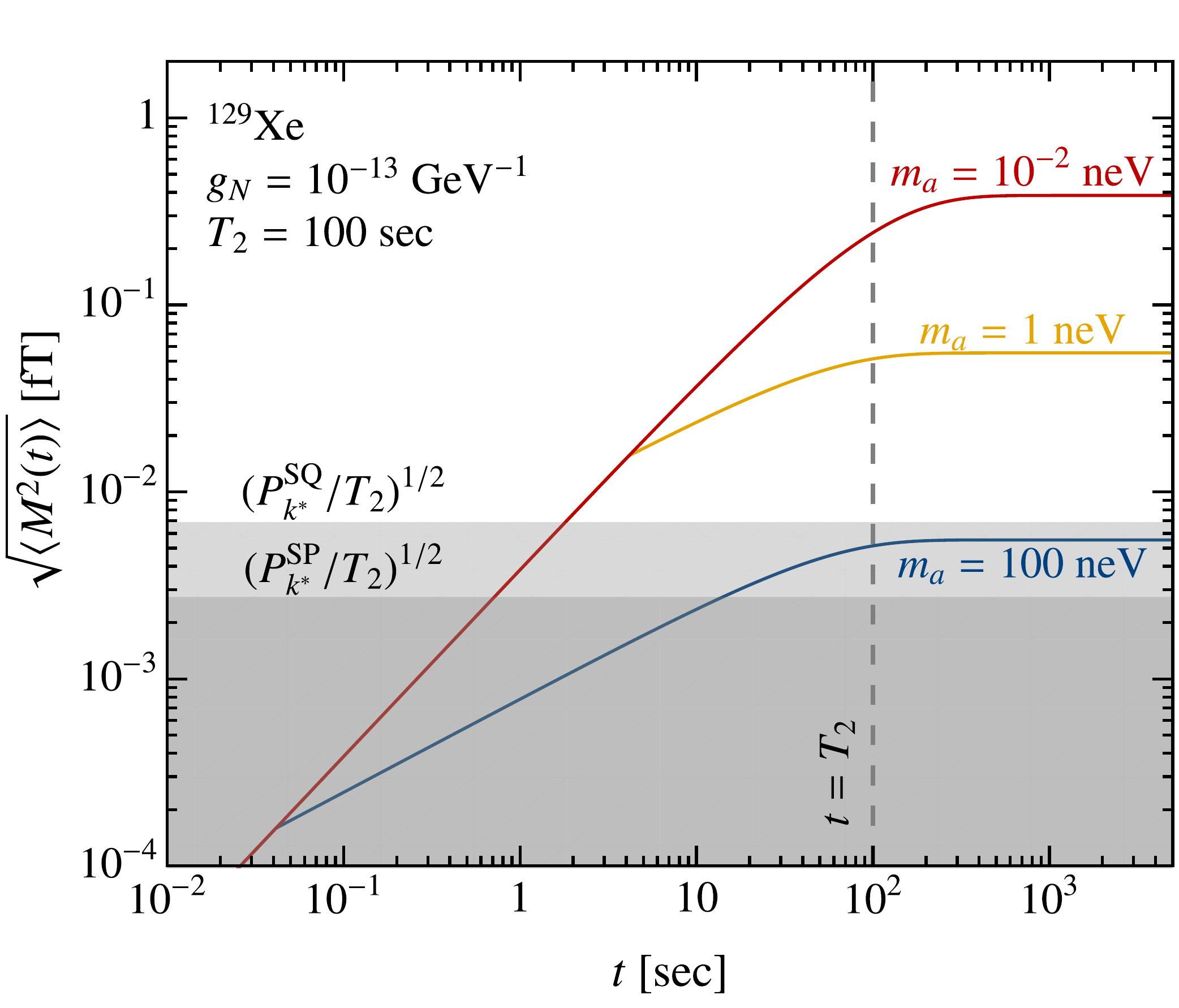} 
\hspace{0.25cm}
\includegraphics[width=0.485\textwidth]{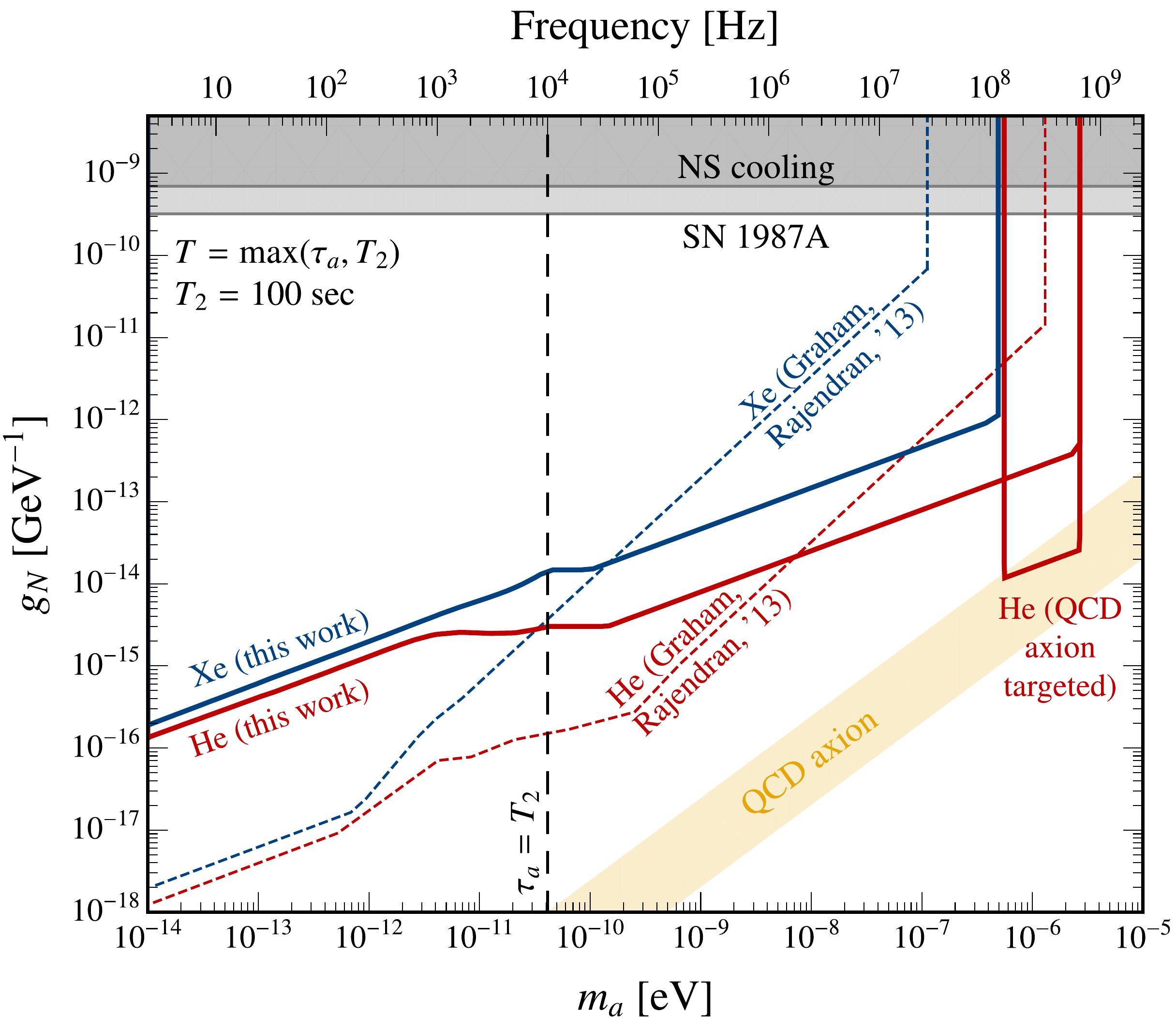} 
\end{center}
\vspace{-0.7cm}
\caption{Projections for the sensitivity of a prototypical spin-precession experiment to the magnetic dipole moment operator, $g_N (\partial_\mu a) \bar{N} \gamma^\mu \gamma_5 N$.
{\bf Left}: The axion-induced growth of the magnetization determined from the Bloch equations.
At short times the magnetization grows $\propto t$, saturating at $T_2$.
If $\tau_a < T_2$, the amplitude grows $\propto \sqrt{t}$ for $\tau_a < t < T_2$.
The magnitudes are compared to those from a SQUID and spin-projection noise, evaluated at $k^* \equiv \omega_0 T/2\pi$, and assuming an integration time $T=T_2$.
{\bf Right}: Projected sensitivity to $g_N$ (solid) in comparison to previous projections of \Refc{Graham:2013gfa} (dashed), for an almost identical scan strategy.
Strengthened sensitivity for $m_a \gg 2\pi/v^2 T_2$ arises primarily due to the signal growth we account for when $\tau_a < t < T_2$, whereas the suppression at lower masses arises partially from a refined estimate of the spin-projection noise.
Blue and red lines correspond to Xenon and Helium targets, and for the latter we label a ``QCD axion targeted'' for the results that could be obtained with negligible spin-projection noise and five years of integration time.
}
\label{fig:sensitivity}
\vspace{-0.5cm}
\end{figure*}

Eqs.~\eqref{eq:MMcont} and~\eqref{eq:MMstoc} allow us to infer the growth of the magnetization in the presence of a finite $\tau_a$, as $C(t,t) = \langle M_x^2(t) \rangle$.
Firstly, for $t \gg T_2$, we see that the growth saturates even with a finite coherence time in the driving force, implying saturation occurs in this limit regardless of the size of $\tau_a$.
For $t \ll T_2$, however, the behaviors differ,
\begin{equation}
\lim_{t \ll T_2} \langle M_x^2(t) \rangle = \frac{\langle A^2 \rangle}{8 \omega_0^2} 
\left\{\begin{array}{lc} t^2 & t \ll \tau_a, \\
\tau_a t & t \gg \tau_a.
\end{array} \right.
\label{eq:twogrowths}
\end{equation}
The first result, that the amplitude of the magnetization grows linearly with time for $t \ll \tau_a,\,T_2$, is consistent with \Eq{eq:Mx}.
However, for $\tau_a < t < T_2$, we see that the amplitude continues to grow, leading to an ever-increasing size, albeit as $\sqrt{t}$.
Intuitively, this transition in behavior can be understood as the magnetization executing a random walk.
For $t > \tau_a$, $M_x(t)$ in Eq.~\eqref{eq:Mxsol} is now a sum of contributions from the axion field at different coherence times, all of which are out of phase.
The sum is analogous to a 2D random walk with steps of length $\tau_a$ (given the growth for $t < \tau_a$), and with a number of steps $t/\tau_a$, so that we expect $|M_x(t)| \propto \sqrt{t \tau_a}$, exactly as found (cf. \Refc{Foster:2017hbq}).
In \Fig{fig:sensitivity} ({\bf Left}), we show the growth of the magnetization for various axion masses, computed directly from \Eq{eq:int}.
The three regimes ($t < \tau_a$, $\tau_a < t < T_2$, $T_2 < t$) can be clearly observed.

To compute the experimental sensitivity to a highly coherent axion signal, it is convenient to move to the frequency domain.
We imagine a dataset $\{M_n = M_x(n \Delta t)\}$ of measurements of the magnetization collected at a frequency $1/\Delta t$, for an integration time $T = N \Delta t$.
We can then compute the power spectral density (PSD) as~\footnote{In Eq.~(\ref{eq:PSDdef}) we specify the expectation value of the PSD, rather than the PSD itself.
Given a single experimental dataset, the average PSD cannot be computed, and it will vary between realizations as the axion field which gave rise to it is stochastic.
Nevertheless, our goal is to determine what value we expect to generate, and so we will compute the expected PSD throughout.
A more careful distinction is provided in the SM},
\begin{equation} 
P_k = \frac{\Delta t^2}{T} \langle|\tilde{M}_k|^2 \rangle,\hspace{0.5cm} 
\tilde{M}_k \equiv \sum_{n = 0}^{N-1} e^{-i2\pi k n/N} M_n.
\label{eq:PSDdef}
\end{equation} 
We can compute the PSD exactly for arbitrary $T$, $T_2$, and $\tau_a$~\footnote{This is discussed in detail in the SM.
Also see \cite{Lisanti:2021vij,Gramolin:2021mqv,Lee:2022vvb,Gao:2022nuq} for related considerations.}.
The result is particularly simple for $T \gg T_2$, taking the form
\begin{equation} 
P_k^a \simeq \frac{\langle A^2 \rangle T_2^2}{16 \omega_0^2} 
\left\{ \begin{array}{lc} 
\displaystyle\frac{4}{\Delta \omega^2_k T} \sin^2 \left[ \tfrac{1}{2} \Delta \omega_k T \right] & T \ll \tau_a, \\[0.4cm]
\displaystyle\frac{\tau_a}{ 1 + \Delta \omega^2_k T_2^2} & T \gg \tau_a,
\end{array} \right.
\label{eq:Pk}
\end{equation}
with $\Delta \omega_k \equiv 2\pi k / T - \omega_0$.
Due to the resonant response of the sample, the signal peaks for $\Delta \omega_k \simeq 0$.
For $T_2 \ll T \ll \tau_a$, the signal falls dominantly in a single $k$-bin, or exactly if $\omega_0 T/2\pi \in \mathbb{N}$.
Once $T > \tau_a,\, T_2$, the signal becomes resolved into multiple bins.
This will impact the signal-to-noise scaling, as we will discuss after considering the relevant background contributions.

\vspace{0.2cm}
\noindent {\bf Noise Sources.}
%
The axion signal must be detected on top of three relevant Gaussian background contributions: thermal noise, SQUID noise, and spin-projection noise~\footnote{In both the fluctuating phase and plane wave models, the axion field itself forms a Gaussian random field, and hence one can show so too does any linear functional of the axion (see, e.g., Ref.~\cite{Weinberg:2008zzc}, Appendix E).}.
Thermal noise arises in the readout circuit and can be suppressed by cooling the apparatus, and we assume this can be done sufficiently for this noise source to be neglected (for details, see \Refc{Aybas:2021cdk}).
The transverse magnetic field, related to the transverse magnetization by an ${\cal O}(1)$ factor which we take to be unity (see the SM), is read out using a SQUID.
This magnetometer noise is frequency-independent for $f \gtrsim 10~{\rm Hz}$~\cite{PhysRevLett.110.147002,Kahn:2016aff} and can be modeled as,
\begin{equation}
C_{\rm SQ}(t,t') = \delta(t-t') \frac{1}{A_{\rm eff}^2} P_{\Phi\Phi}^{\rm SQ}  
\hspace{0.1cm}\Rightarrow\hspace{0.1cm}
P_k^{\rm SQ} = \frac{1}{A_{\rm eff}^2} P_{\Phi\Phi}^{\rm SQ},
\label{eq:PSQ}
\end{equation}
where by default, we take $P_{\Phi \Phi}^{\rm SQ} \simeq (\mu\Phi_0)^2/{\rm Hz}$, with $\Phi_0$ the magnetic flux quantum, and $A_{\rm eff}$ is the effective area of the sample sensed by a pickup loop, which we take to be $\simeq 0.3~{\rm cm}^2$, following \Refc{Budker:2013hfa}.

The final background source we consider is spin-projection noise, which originates directly from the quantum nature of the nuclear spins in the sample~\cite{PhysRevB.75.085310}.
We can determine its magnitude from the autocorrelation function.
Consider two successive measurements of the spin operator along the $\hat{\bf x}$ direction, $ S _x $, taken at $t$ and then $t'$.
The operators are related through the time evolution operator, $\mathcal{U}(t) = \exp(i\omega_0 S_z t)$, $S_x(t') = \mathcal{U}^\dag(t'-t) S_x(t)\, \mathcal{U}(t'-t).$
The magnetization can be determined from the sum over all nuclear spins, and assuming the sample is hyperpolarized (i.e. unit polarization fraction), we obtain,
\begin{align}
C_{\rm SP} (t,t') &= \frac{\gamma^2}{2V^2} e^{-|t-t'|/T_2} \sum_i \big\langle S_x^{(i)}(t) S_x^{(i)} (t') \big\rangle + {\rm h.c.}  \nonumber\\ 
& = \frac{\gamma^2nJ}{2V} e^{-|t-t'|/T_2} \cos \left[ \omega_0 (t-t') \right]\!,
\label{eq:CSP}
\end{align} 
where $J$ is the nuclear spin, $n$ is the number density of spins, and $V$ is the volume of the sample.
The exponential factor is included to account for transverse-spin relaxation, and for further details, see the SM.
\Eq{eq:CSP} exhibits a $V^{-1}$ scaling, which suggests that for a large enough sample, this noise source can be suppressed.
The corresponding PSD for $t,t' \gg T_2$ is given by,
\begin{equation} 
P_k^{\rm SP} = \frac{\gamma^2 n J}{2V} \frac{T_2}{1 + \Delta \omega^2_k T_2^2}\,.
\label{eq:Pkspin}
\end{equation} 
This result determines the spin-projection noise for an arbitrary $J$, assuming a hyperpolarized sample.

\vspace{0.2cm}
\noindent {\bf Experimental Sensitivity.}
%
We now combine the above results to forecast the expected sensitivity to an axion-induced signal. We use the signal PSD in the $\tau _a \gg T$ and $\tau _a \ll T$ limits and compare it to the background PSD, using all $k$-bins and the likelihood framework presented in the SM (employing the formalism of \Refc{Foster:2017hbq} and inserting the Asimov dataset~\cite{Cowan:2010js}).
Above the transition region of $\tau _a = T$, we interpolate between the two regimes with a horizontal line. In principle, one could extend our analysis to handle the intermediate regime.
Detailed projections require us to specify explicit experimental parameters.
Even before this, we can determine the sensitivity scaling with integration time, which varies depending on the hierarchy between $\tau_a$, $T_2$, and $T$.
Specifically,
\begin{equation}\begin{aligned}
T \ll \tau_a,\,T_2  &\hspace{0.2cm}\Rightarrow\hspace{0.2cm} g \propto T^{-3/2} \\ 
\tau_a \ll T \ll T_2 &\hspace{0.2cm}\Rightarrow\hspace{0.2cm} g \propto T^{-1} \\
T_2 \ll T \ll \tau_a &\hspace{0.2cm}\Rightarrow\hspace{0.2cm} g \propto T^{-1/2} \\
\tau_a,\,T_2 \ll T &\hspace{0.2cm}\Rightarrow\hspace{0.2cm} g \propto T^{-1/4},
\label{eq:paramgrowth}
\end{aligned}\end{equation}
where $g=g_N,g_d$.
The growth becomes slowest once $T$ is the largest timescale, and the signal is resolved into multiple bins.
These scalings are derived in the SM, however, they arise from comparing the growth of the signal and background.
For instance, for $T_2 \ll T \ll \tau_a$, where the signal is dominantly in a single bin, $k^* \equiv \omega^0 T/2\pi$, we see that $P_{k^*}^a \propto T$ from \Eq{eq:Pk}, whereas $P_{k^*}^{\rm SQ}$ and $P_{k^*}^{\rm SP}$ are independent of $T$.
Estimating sensitivity by matching the signal to the background, we find the limit scales as $g \propto T^{-1/2}$.

To provide quantitative projections, we match the parameters specified in the CASPEr papers~\cite{Graham:2011qk,Budker:2013hfa}, focusing on the magnetic dipole operator.
The accessible Larmor frequencies set the mass range we consider, and we take $10^{-14}~{\rm eV} < \omega_0 < \gamma B_{\rm max}$, with $B_{\rm max}$ the maximum magnetic field.
We fix $T_2 = 100~{\rm sec}$ and for each mass, adopt a variable integration time, $T = {\rm max}[\tau _a , T_2]$, to ensure we always run until the $T^{-1/4}$ growth sets in from \Eq{eq:paramgrowth}.
This implies that the signal, and therefore the likelihood, remains dominated by a single frequency bin.
To ensure each mass is covered only once, resonant frequencies are adjusted by $2\pi\, {\rm max}[\tau_a^{-1},T_2^{-1}]$.

We consider two different spin-1/2 samples, and both assumed to be hyperpolarized.
The first consists of pure Xenon-129, which has a nuclear magnetic moment of $0.78~\mu_N$ and a nuclear spin density of $1.3 \times 10^{22}~{\rm cm}^{-3}$, and we assume $B_{\rm max} = 10~{\rm T}$.
The second is a more optimistic projection using Helium-3 -- $\mu = 2.12~\mu_N$ and $n = 2.8 \times 10^{22}~{\rm cm}^{-3}$ -- and $B_{\rm max}=20~{\rm T}$, as well as assuming the SQUID noise can be decreased by two orders of magnitude below that discussed around \Eq{eq:PSQ}.
As shown in the SM, to cover the full mass range, these two experiments would require a total integration time of 56.1 and 61.5 years, respectively.

Our results are shown in \Fig{fig:sensitivity} ({\bf Right}), and contrasted with the projections of \Refc{Graham:2013gfa}.
The discrepancy has at least three sources: 1. the additional growth the signal we have accounted for when $\tau_a < T < T_2$, see \Eq{eq:twogrowths}; 2. a different treatment of the spin-projection noise; and 3. a different calculation of the Larmor frequency given $B_{\rm max}$.~\footnote{For Helium our result is exactly a factor of two larger, whereas for Xenon the difference is slightly larger than a factor of 4.
For Xenon, this difference is attributed to the use of a combination of isotopes, whereas we considered a pure sample of Xenon-129.} 
For the spin-projection noise, we compute the value for each $k$ with \Eq{eq:Pkspin} -- recall the signal is dominantly peaked in a single bin -- rather than integrating a result over a range of frequencies near $\omega_0$, as in Eq.~(A2) in \Refc{Budker:2013hfa}.
Further discussion is provided in the SM, where we also contrast our results for the electric dipole operator.
As a final benchmark, in \Fig{fig:sensitivity}, we also show the Helium-3 sensitivity assuming that spin-projection noise could be evaded, and the mass range is set by assuming five years of integration time.
This benchmark cuts into the QCD axion parameter space, illustrated by the yellow band \footnote{The band is defined as the region bounded from below by the KSVZ axion model and from above by the DFSZ model with the down-type Higgs doublet vacuum expectation value set to zero.
(Note that the full range of DFSZ Higgs' vacuum expectation values allows the QCD axion parameter space to fill the entire lower right corner of \Fig{fig:sensitivity}.)}.

\vspace{0.2cm}
\noindent {\bf Discussion.}
%
In this work we have derived the axion-induced signal in spin-precession experiments.
These instruments remain one of the most promising paths to measuring axion-induced magnetic and electric dipoles, and our calculations demonstrate that their sensitivity is significantly different to what has previously been estimated.
Arguably our most important finding is the enhanced detection prospects for the QCD axion at high masses, a result which arises from the continued growth of the axion-induced signal when integrating beyond the axion coherence time \footnote{In light of our results, one may wonder if its possible to build an experiment using an element that carries a larger gyromagnetic ratio than Helium-3 to increase the maximum testable axion mass and improve the coverage of the QCD axion line.
Surveying all measured gyromagnetic moments of nuclear isotopes, the only element with a substantially larger value is Thallium-200, although since it carries a half-life of $\sim$1 day it does not appear viable.
Tritium, atomic Hydrogen, and Fluorine-19 have larger gyromagnetic ratios than Helium-3, but by less than 40\%.}.
Our findings have broader implications.
To name one, they demonstrate that signals that are less coherent than dark matter -- for instance, a Cosmic axion Background~\cite{Dror:2021nyr} -- are more detectable with spin-precession instruments than may otherwise have been concluded~\cite{CaB-Nuclear:upcoming}. \nocite{Gramolin:2021mqv}

\vspace{0.5cm}
\noindent {\it Acknowledgements.}
%
Our work benefited from conversations with Hendrik Bekker, Yonatan Kahn, Alexander Sushkov, and Arne Wickenbrock.
Further, we thank the anonymous referee for useful feedback.
J.M.L. is supported by the Deutsche Forschungsgemeinschaft under Germany's Excellence Strategy - EXC 2121 ``Quantum Universe'' - 390833306.
The research of JD and SG is supported in part by NSF CAREER grant PHY-1915852 and in part by the U.S. Department of Energy grant number DE-SC0023093.
Part of this work was performed at the Aspen Center for Physics, which is supported by National Science Foundation grant PHY-1607611.

\bibliographystyle{JHEP}
\bibliography{NMR_Refs}

\clearpage
\onecolumngrid
\begin{center}
   \textbf{\large SUPPLEMENTARY MATERIAL \\[.2cm] ``On the Sensitivity of Spin-Precession Axion Experiments''}\\[.2cm]
  \vspace{0.05in}
  {Jeff A. Dror, Stefania Gori, Jacob M. Leedom, and Nicholas L. Rodd}
\end{center}
\setcounter{equation}{0}
\setcounter{figure}{0}
\setcounter{table}{0}
\setcounter{page}{1}
\setcounter{section}{0}
\makeatletter
\renewcommand{\thesection}{S-\Roman{section}}
\renewcommand{\theequation}{S-\arabic{equation}}
\renewcommand{\thefigure}{S-\arabic{figure}}

\section{Electric Dipole Operator}
\label{app:EDM}

In the main text, we focused on the impact of our results on the magnetic dipole operator in \Eq{eq:ops}.
To measure the electric dipole operator, \Refs{Graham:2013gfa,Budker:2013hfa} proposed using a hyperpolarized sampled with an additional applied electric field.
Note that as the electrons adjust themselves to shield the field, the nucleus experiences a smaller effective field ${\bf E}^\ast$.
To minimize the shielding, one needs to carefully choose the sample and \Refs{Graham:2013gfa,Budker:2013hfa} proposed Lead-207.
While there are significant differences in the instrument required to measure the electric dipole operator, in terms of the calculation we presented in the main text, the primary difference is the $A$ coefficient controlling the magnitude of the driving force (defined in \Eq{eq:F}).
Accordingly, we can readily extend our results to this operator, simply by making the replacement $g _N \to \sqrt{3} g_d E^\ast/\omega_0 v$, see \Eq{eq:Asq}.

The experiment proposed in Refs.~\cite{Graham:2013gfa,Budker:2013hfa} has two phases, with both utilizing a lead sample of diameter $10~{\rm cm}$ and $E^\ast = 3\times 10^{10}~{\rm V/m}$.
In the first phase (P1), the polarization $p$, relaxation time $T_2$, and maximum magnetic field $B_{\rm max}$ are taken to be $10^{-3}$, $1~{\rm msec}$, and $10~{\rm T}$, respectively.
In the second phase (P2), these same parameters are given values $p=1$, $T_2=1~{\rm sec}$, and $B_{\rm max}=20~{\rm T}$.
For each phase, we envision scanning from $10^{-14}~{\rm eV}$ up to the maximum accessible Larmor frequency, and as discussed in App.~\ref{app:inttime}, the total scan time is 5.41 hours (213 days) for P1 (P2).

\begin{figure*}[!t]
\begin{center} 
\includegraphics[width=0.65\textwidth]{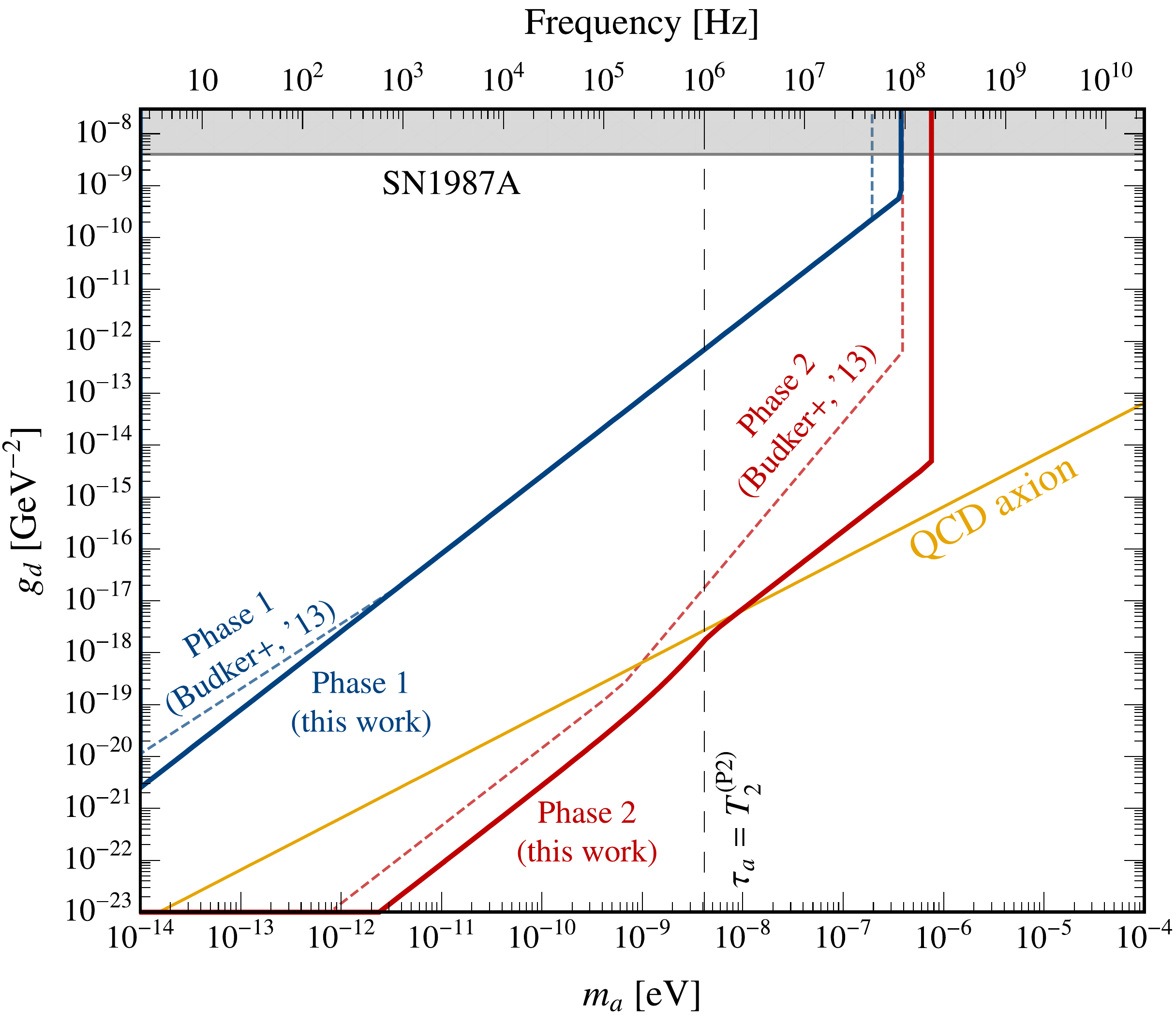} 
\end{center}
\vspace{-0.7cm}
\caption{Projected sensitivity for the electric dipole operator.
The experimental proposal has two phases corresponding to the different parameters discussed in the text. For Phase 1, the experiment never reaches the transition for the stochastic regime while for Phase 2, the transition region is shown by the vertical dashed line.
For this operator the QCD axion line is well predicted, and forms a strongly motivated target.
The bound from SN1987A is taken from Ref.~\cite{Graham:2013gfa}.}
\label{fig:EDM}
\vspace{-0.5cm}
\end{figure*}

Using the expressions provided in the main text, we compute the expected sensitivity shown as the solid lines in \Fig{fig:EDM}.
The different scaling between the mass and coupling in \Fig{fig:sensitivity} ({\bf Right}) and \Fig{fig:EDM} arises from the appearance of $\omega_0 \simeq m_a$ in the amplitude of the driving force for the magnetic dipole operator, see \Eq{eq:F}.
More importantly, we again see clear differences to the projections from the earlier results, here taken from \Refc{Budker:2013hfa}.
Firstly, we find that the maximum mass that can be obtained is a factor of two larger than the earlier calculation.
Next, note that the spin-projection noise for Lead-207 is subdominant to the background from the SQUID, and so our different treatment of $P_k^{\rm SP}$ has no impact in this figure (cf. \Fig{fig:sensitivity}).
For P1, aside from the maximum mass, we find very good agreement.
This arises as for $T_2 = 1~{\rm msec}$, $T_2 < \tau_a$ across the entire mass range.
Therefore, at no point can the magnetization enter the additional period of growth we predict for $\tau_a < t < T_2$.
For P2, $\tau_a < T_2$ at higher axion masses, implying a range where we predict additional growth, and indeed our sensitivity projection begins to parametrically separate from the earlier result.

\section{Spin-Projection Noise}
\label{app:SPN}

In this appendix, we derive the expression for the spin-projection noise given in Eqs.~\eqref{eq:CSP} and \eqref{eq:Pkspin} in the main text.
The aim is to determine the autocorrelation of spins measured in the $x$-direction, given a sample with $N_s$ nuclei of spin $J$ and gyromagnetic ratio $\gamma$, in a uniform magnetic field ${\bf B}_0 = B_0 \hat{\bf z}$.
The operator for the total spin of the system is the sum of the individual nuclear spins,
\begin{equation}
S_z = \sum_{i=1}^{N_s} S_z^{(i)}.
\end{equation}
For each nuclei, we use a basis spanned by the total spin and the projection along the $z$-axis, so that
\begin{equation}
S_z^{(i)}\ket{i;J,m} = m\ket{i;J,m},\hspace{0.5cm}
m=-J, - J + 1 , \ldots ,J.
\end{equation}

Neglecting interactions between nuclei, the Hamiltonian of the system is given by, $H = -\boldsymbol{\mu} \cdot {\bf B} _0  = -\omega_0 S_z$, leading to a time evolution operator, $\mathcal{U}(t) =\exp[i\omega_0S_z t]$.
We now follow \Refc{Graham:2013gfa} and assume that the system is hyperpolarized such that all sites are in the highest $S_z^{(i)}$ eigenstate with $m = J$. The state of the sample is
\begin{equation}
\ket{\uparrow} \equiv \bigotimes_{i=1}^{N_s}\ket{i;J,J}\!.
\end{equation}
First, we compute the autocorrelation for a single spin in the sample, finding,
\begin{equation}
\bra{i;J,J} S^{(i)}_x(t)S^{(i)}_x(t')\ket{i;J,J} = \frac{1}{2} J e^{-i\omega_0(t-t')}.
\end{equation}
The result for the full system follows,
\begin{equation}
\bra{\uparrow} S_x(t)S_x(t^\prime) \ket{\uparrow} 
= \sum_{i=1}^{N_s} \bra{i;J,J} S^{(i)}_x(t) S^{(i)}_x(t') \ket{i;J,J}
= \frac{1}{2} N_s J e^{-i\omega_0(t-t')}.
\end{equation}

We can now derive Eqs.~\eqref{eq:CSP} and \eqref{eq:Pkspin}.
Firstly, the spin-projection autocorrelation function is,
\begin{equation}
C_{\rm SP}(t,t') = \frac{\gamma^2}{2V^2}e^{-|t-t'|/T_2} \bra{\uparrow} S_x(t)S_x(t')+S_x(t')S_x(t) \ket{\uparrow} 
=\frac{\gamma^2n J}{2V}e^{-|t-t'|/T_2} \cos\left[\omega_0(t-t')\right]\,.
\end{equation}
As discussed in the main text, the exponential factor involving $T_2$ is phenomenological and added to account for the transverse spin relaxation.
From here, we can directly compute the PSD, which in the limit of $t,t'\gg T_2$ is,
\begin{equation} 
P_k^{\rm SP} = \frac{\gamma^2 n J}{2V} \frac{T_2}{1 + \Delta \omega^2_k T_2^2}\,, 
\end{equation} 
as shown in the main text.
For $J=1/2$ this agrees parametrically with existing literature~\cite{PhysRevB.75.085310,Graham:2013gfa,Budker:2013hfa,Aybas:2021cdk}.

\section{Power Spectral Densities}
\label{app:PSD}

In the main text, we presented expressions for the autocorrelation functions for the magnetization induced by the axion, as well as for several noise sources.
Nevertheless, our eventual sensitivities were determined in the frequency domain, and in particular, we made use of the PSD defined in \Eq{eq:PSDdef}, as discussed in App.~\ref{app:likelihood}.
Here we briefly expand on the procedure used to compute $P_k$.

To begin with, we assume that $\Delta t$ is sufficiently finely spaced compared to the scale over which $M_x(t)$ varies that we can approximate the sums in the discrete Fourier transform with the following integrals,
\begin{equation} 
P_k \simeq \frac{1}{T} \int_0^T dt \int_0^T dt' e^{i\omega_k (t-t')} C(t,t'),
\end{equation} 
where the angular frequency for a given $k$ mode is $\omega_k \equiv 2\pi k/T$.
To evaluate the integrals, we switch variables to $x = (t-t')/2$ and $y = (t+t')/2$.
Additionally, except for the SQUID noise, the other expressions for $C(t,t')$ that we studied are of the form $f(x,y) \cos [2 \omega _0 x]$, allowing us to write,
\begin{equation}
P_k = \frac{2}{T} \int_0^{T/2} dx  \cos \left[ 2 (\omega_0 - \omega_k) x \right] \int_x^{T-x} dy f(x, y),
\end{equation}
where we have dropped a rapidly oscillating term and exploited $f(x,y) = f(-x,y)$.

Following the procedure above, we can exactly evaluate $P_k$ for each expression we consider, and we use these expressions when computing the limits we show in Figs.~\ref{fig:sensitivity} and \ref{fig:EDM}.
If we assume $\tau_a,\,T,\, T_2 \gg \omega_0^{-1}$, which holds for almost our entire parameter space, the expressions simplify, and we have
\begin{align}
\lim_{T \ll \tau_a} P _k^a &= \frac{\langle A^2 \rangle T_2^2}{16 \omega_0^2} \frac{1}{\Delta \omega_k^2 T} \frac{1}{1 + \Delta \omega_k^2 T_2^2} \left[ 4 \sin^2 \left[\tfrac{1}{2} \Delta \omega_k T\right]  + \left(\sin [\Delta \omega_k T] - \left[ 1 - e^{-T/T_2} \right] \Delta \omega_k  T_2 \right) ^2 - \sin^2 [\Delta \omega_k T] \right]\!,\nonumber\\[0.2cm]
\lim_{T \gg \tau_a} P_k^a &= \frac{\langle A^2 \rangle T_2^2}{16 \omega_0^2} \frac{\tau_a}{1 + \Delta \omega^2_k T_2^2}\left[1 - \frac{T_2}{T} \left( \frac{1}{2} \left[1 + e ^{-2T/T_2} \right] + \frac{1-\Delta\omega_k^2 T_2^2}{1 + \Delta \omega_k^2 T_2^2} \right.\right. 
\label{eq:Pkafull}\\
&\hspace{4cm}\left.\left.- 2 e^{-T/T_2} \frac{\cos [\Delta \omega_k T] - \Delta \omega_k T_2 \sin [\Delta \omega_k T]}{1 + \Delta \omega^2_k T^2_2} \right) \right]\!, \nonumber\\[0.2cm] 
P_k^{\rm SP} &= \frac{\gamma^2 n J}{2V}  \frac{T_2}{1 + \Delta \omega_k^2 T_2^2}  \left[1 - \frac{T_2}{T} \left( \frac{1 - \Delta \omega_k^2 T_2^2}{1 + \Delta \omega_k^2 T_2^2} - e^{-T/T_2} \frac{(1-\Delta \omega_k^2 T_2^2) \cos [\Delta \omega_k T] - 2 \Delta \omega_k T_2 \sin [\Delta \omega_k T]}{1 + \Delta \omega_k^2 T_2^2} \right) \right]\!. \nonumber
\end{align}
In the limit where $T \gg T_2$, these expressions reduce to those in Eqs.~\eqref{eq:Pk} and \eqref{eq:Pkspin}, and in App.~\ref{app:likelihood}, we will use these to confirm the scalings in \Eq{eq:paramgrowth}.
We have not restated $P_k^{\rm SQ}$, as the exact expression was already given in \Eq{eq:PSQ}.

\section{Likelihood Framework}
\label{app:likelihood}

Here we expand on the likelihood framework used to compute the projected limits throughout this work.
The formalism we employ is similar to that of \Refc{Foster:2017hbq}, and we refer there for additional details.
The PSD for each contribution we consider is exponentially distributed, and therefore the exponential distribution provides the appropriate likelihood.
Our dataset, $d$, is given by the time series of magnetization measurements, $M(n \Delta t)$, $n = 0,1,\ldots,N-1$, taken over the integration time $T = N \Delta t$.
From the experimentally measured values, we can compute the PSD for the data in each bin, $P_k^{(d)}$.
If we then model the PSD in each bin by $P_k$, the likelihood can be written as
\begin{equation}
{\cal L}(d\, | P_k) = \sum_{k=1}^{N-1} \frac{1}{P_k} e^{-P_k^{(d)}/P_k}.
\end{equation}

We will decompose our PSD model into a signal and background component as $P_k \equiv S_k + B_k$.
The result for the signal is $P_k^a$, as given in \Eq{eq:Pk}, whereas for the background, we combine the SQUID and spin-projection noises as $B_k = P_k^{\rm SQ} + P_k^{\rm SP}$, using Eqs.~\eqref{eq:PSQ} and \eqref{eq:Pkspin}.
In order to determine the projected 95\% exclusion limits, we will use a log-likelihood ratio test statistic, and further make use of the Asimov dataset~\cite{Cowan:2010js} rather than generating Monte Carlo simulations.
In particular, we compute
\begin{equation}
q = 2 \sum_{k=1}^{N-1} \left[ \left( 1 - \frac{B_k}{S_k+B_k} \right) - \ln \left( 1 + \frac{S_k}{B_k} \right) \right]\!,
\label{eq:TSq}
\end{equation}
and then determine the amplitude of the signal where $q=-2.71$, which will correspond to the limiting value.

When computing estimated limits, we make use of the full expression in \Eq{eq:TSq}.
However, we can gain intuition from approximate versions of this result.
To begin with, as noted in the main body, until $T$ is larger than $T_2$ and $\tau_a$, $S_k$ will be non-zero in all bins except $k^* \simeq \omega_0 T/2\pi$.
This is the case for our scan strategy when we never allow $T$ to become parametrically larger.
Accordingly, only a single bin in \eqref{eq:TSq} contributes, and the 95\% limit occurs when
\begin{equation}
\frac{S_{k^*}}{B_{k^*}} = - \frac{1}{W_0(-e^{-2.71/2-1})} - 1 \equiv \kappa.	
\label{eq:Lim-onebin}
\end{equation}
Here $W_0$ is the principal branch of the Lambert $W$ function, and from this we can compute $\kappa \simeq 8.48$.
Note this implies that in the single bin limit, the properties of the exponential distribution imply that we can only exclude a value of the signal noticeably larger than the background.
To probe $S_k \ll B_k$, we will need multiple bins, and therefore $T \gg T_2,\tau_a$, a case we return to below.

From \Eq{eq:Lim-onebin}, we see that the 95\% limit is obtained for $S_{k^*} \simeq \kappa B_{k^*}$ whenever $T$ is not the largest timescale.
We can then use this result to confirm the first three scalings in \Eq{eq:paramgrowth}.
In particular, from \Eq{eq:Pkafull},
\begin{equation} 
P_{k ^\ast }^a \simeq \frac{\langle A^2 \rangle}{16 \omega_0^2} 
\left\{ \begin{array}{lc} 
\displaystyle\tfrac{1}{4} T^3 & T \ll \tau_a,\,T_2, \\[0.3cm]
\displaystyle\tfrac{1}{3} T^2\, \tau_a & \tau_a \ll T \ll T_2, \\[0.3cm]
\displaystyle T\, T_2^2 & T_2 \ll T \ll \tau_a, \\[0.3cm]
\displaystyle \tau_a\, T_2^2 & \tau_a,\,T_2 \ll T,
\end{array} \right.
\end{equation}
As $\langle A^2 \rangle \propto g^2$, the claimed scalings follow immediately from the first three results above.

Although we did not consider this case in our scan strategy, let us take $T \gg \tau_a,\,T_2$, so that the signal becomes resolved into many bins.
From the above, we see that $P_k^a$ is no longer growing with $T$ in this limit.
As larger values of $T$ correspond to a finer frequency spacing of the $k$ values in the PSD, for large enough $T$, we can replace the sum in \Eq{eq:TSq} with an integral,
\begin{equation}
q \simeq \frac{T}{\pi} \int_0^{\infty} d\omega \left[ \left( 1 - \frac{B(\omega)}{S(\omega)+B(\omega)} \right) - \ln \left( 1 + \frac{S(\omega)}{B(\omega)} \right) \right]\!,
\label{eq:qint}
\end{equation}
where $S(\omega) = S_{k=\omega T/2\pi}$, and similarly for $B(\omega)$.
We note that, in actuality, the frequency integration will extend to $2 \pi (N-1)/T$ rather than $\infty$. As the signal we consider will be strongly peaked near the resonant frequency, this distinction is not important.
Further, we note that this same approach was used by default in \Refc{Foster:2017hbq}, as instruments searching for the axion-photon coupling commonly work in the regime where $T$ is the largest timescale.
If we assume that we can now probe $S(\omega) \ll B(\omega)$, then the limit is set when
\begin{equation}
q \simeq -\frac{T}{2\pi} \int_0^{\infty} d\omega\, \frac{S^2(\omega)}{B^2(\omega)}.
\end{equation}
In particular, as $S^2(\omega) \propto g^4$, we see that in this limit, the sensitivity is scaling as $g \propto T^{-1/4}$, as claimed in \Eq{eq:paramgrowth}.

As a final comment, we note that sensitivity to $S \ll B$ -- and therefore the validity of \Eq{eq:qint} -- requires a large number of bins and also $T$ to be parametrically larger than any other timescale.
We can demonstrate this with a simple example.
Consider a signal present in $n_S$ bins, with a PSD value of $S$ in each, and similarly assume a flat background PSD, $B$.
Using \Eq{eq:TSq}, we find the contribution to $q$ is the same from every bin, so in direct analogy to \Eq{eq:Lim-onebin} we obtain the following sensitivity
\begin{equation}
\frac{S}{B} = - \frac{1}{W_0(-e^{-2.71/2n_S-1})} - 1.
\end{equation}
For $n_S = 1$, we recover the single bin result of $S/B = 8.48$.
To obtain $S/B = 1$ requires $n_S \simeq 7$, whereas $S/B = 0.1$ occurs for $n_S \simeq 300$.
For \Eq{eq:qint} to be percent level accurate requires over $1,000$ bins.

\section{Plane Wave Model}
\label{app:planewave}

In this section, we frame the discussion from the main text in the language of the plane wave model for axion dark matter, thereby providing an alternative to the fluctuating phase model used in the main body.
All results can be restated in this alternative approach, however, we focus on the magnetic dipole operator for simplicity.
In particular, we obtain the same result for the magnetization PSD induced by the axion in the limit $\tau_a \ll T$, demonstrating both models can account for the finite axion coherence time.
In this plane wave model, motivated by the macroscopic occupation number of axions locally, we treat the dark-matter wave as a sum of $N_a$ plane waves~\cite{Foster:2017hbq,Foster:2020fln,Dror:2021nyr},
\begin{equation} 
a(t,{\bf x}) = \frac{a_0}{\sqrt{N_a}} \sum_{i=1}^{N_a} \cos\left[\omega_it - {\bf k}_i\cdot{\bf x} + \phi_i \right]\!.
\label{eq:pwsum}
\end{equation}
For each state in the sum, we evaluate a random velocity ${\bf v}_i$ and phase $\phi_i\in [0,2\pi)$, where the former is drawn from the underlying dark-matter velocity distribution, and used to form the frequency $\omega_i = m_a (1+ |{\bf v}_i|^2/2)$ and wavenumber ${\bf k}_i=m_a{\bf v}_i$.
In this formalism, we can straightforwardly include any velocity distribution, $f_{\bf v}({\bf v}),$ that we choose.
(Note we add the subscript ${\bf v}$ to clearly indicate this is the velocity and not speed distribution.)
For simplicity we will here take an isotropic distribution,
\begin{equation}
f_{\bf v}(\mathbf{v}) = \frac{1}{\pi^{3/2}v^3} e^{-|\mathbf{v}|^2/v^2}\,,
\end{equation}
with $v \sim 10^{-3}$.
For the magnetic dipole operator, the axion driving force arises from the spatial gradient, which we can take to be evaluated at ${\bf x}=0$ and takes the form,
\begin{equation}
{\bf G}(t) \equiv \nabla a(t) =  \frac{a_0}{\sqrt{N_a}} \sum_{i=1}^{N_a} {\bf k}_i \cos \left[\omega_i t + \phi_i \right]\!.
\label{eq:gradpwsum}
\end{equation}

The PSD of ${\bf G}(t)$ has been studied in \Refs{Lisanti:2021vij,Gramolin:2021mqv,Lee:2022vvb,Gao:2022nuq}, and we refer there for further discussion, although we will not require any details from those works here.
We can now use this expression to model the axion driving force in the Bloch equations, and determine the resulting magnetization, as we did for the fluctuating phase model in the main body.
Focusing on the $\hat{\bf x}$ component of ${\bf M}$, if we assume that $\Delta t \to 0$ while $T = N \Delta t$ remains finite, we obtain
\begin{equation}
|\tilde{M}_k|^2 = \frac{4g_N^2M_0^2}{(\omega_0^2-\omega_k^2)^2+4\omega_k^2/T_2^2}\bigg(\omega_0^2|  \tilde{G}^x_k|^2 + \omega_k^2|\tilde{G}^y_k|^2 -2\omega_0\omega_k\text{Im}[\tilde{G}^x_k(\tilde{G}^y_k)^*]\bigg)\,,
\label{eq:master}
\end{equation}
where again $\omega_k = 2\pi k/T$, and $\tilde{G}_k^s$ is the Fourier transform of $\hat{\bf s}\cdot {\bf G}(t)$.
This result encapsulates much of the physics described in the main text.
The transfer function factor multiplying the gradient PSDs has a Lorentzian-like shape with resonant frequency $\omega_0$ and approximate width $T_2^{-1}$; the larger $T_2$, the narrower the response of the sample.
Even when $\tau_a \ll T$, if $T \ll T_2$, then the axion is resolved in the frequency domain, but the magnetization will still lie within a single bin given the narrow sample response.
In the limit where $T\ll \tau_a$, the $\tilde{G}^i_k$ have an essentially Dirac $\delta$-function support in frequency, and the signal again lies in a single bin.

Let us now focus on the high mass parameter space where $\tau_a\ll T=T_2$ and show that the fluctuating phase and plane wave models agree for the signal PSD up to ${\cal O}(1)$ factors.
Assuming an isotropic dark-matter distribution, near $\omega_k\simeq \omega_0$, we have
\begin{equation}
\langle\lvert \tilde{G}_k^x\rvert^2\rangle = \langle\lvert\tilde{G}_k^y\rvert^2\rangle \simeq \frac{2\pi^2}{3} m_a a_0^2 v^3 f_{\bf v}(v),
\end{equation}
and the third term in \Eq{eq:master} averages to zero.
Then we find
\begin{equation}
P_k^a = \frac{4\sqrt{\pi}}{3e} \frac{g_N^2M_0^2 m_a a_0^2 T_2^2}{1+\Delta\omega_k^2T_2^2}.
\end{equation}
This matches the second expression in \Eq{eq:Pk} up to a factor of $\simeq 1.7$.

The plane wave model further provides an alternative explanation for the $\sqrt{t}$ growth exhibited by the magnetization when $\tau_a \ll t \ll T_2$, seen in \Eq{eq:twogrowths}, and underlying our enhanced sensitivity at higher masses seen in \Fig{fig:sensitivity}.
To make the result maximally transparent, we take $T_2 \to \infty$, as it plays no role in the physics of interest.
Further, we adopt a simplified version of the plane wave model, where ${\bf k}_i = m_a v \hat{\bf x}$ for all $i$, so that
\begin{equation}
\nabla a = \frac{a_0 m_a v \hat{\bf x}}{\sqrt{N}} \sum_{i=1}^{N_a} \cos \left[\omega_i t + \phi_i \right]\!.
\end{equation}
If we take $\omega_0 = m_a$, and write $\omega_i = m_a + \delta \omega_i$, with $\delta \omega_i \ll m_a$, the axion-induced magnetization is given by
\begin{equation}
M_x(t) \simeq g_N a_0 m_a v M_0 \frac{1}{\sqrt{N_a}} \sum_{i=1}^{N_a}\, [c_i \sin(m_a t) + s_i \cos (m_a t)] \frac{\sin \left[ \tfrac{1}{2} \delta \omega_i t \right]}{\tfrac{1}{2} \delta \omega_i},
\end{equation}
where $c_i = \cos \phi_i$ and $s_i = \sin \phi_i$.
We see two frequencies that enter the sums, the oscillations at the resonant frequency $m_a$ and the oscillations at the beat frequency $\delta \omega_i$.
Accordingly,
\begin{equation}
\langle M_x^2(t) \rangle \simeq (g_N a_0 m_a v M_0)^2 \frac{1}{2N_a} \sum_{i=1}^{N_a}\, \left\langle \left( \frac{\sin \left[ \tfrac{1}{2} \delta \omega_i t \right]}{\tfrac{1}{2} \delta \omega_i} \right)^2 \right\rangle\!.
\end{equation}
This result exhibits the relevant behavior.
When $t \ll \tau_a$, we cannot resolve the various frequencies of the axion field, and therefore $\delta \omega_i t \ll 1$.
We recover the early $M_x(t) \propto t$ growth in this case.
Once $t \gg \tau_a$, there will be frequencies for which $\delta \omega_i t$ is no longer small, we have passed the timescale of their associated beat.
The magnetization contributed by these terms will turn over, and quickly they will no longer contribute to the sum.
However, if we expect the frequencies are uniformly distributed, there will still be a fraction of frequencies of size $\tau_a/t$ for which $\delta \omega_i t \ll 1$, and which remain in the linear growth regime.
Accordingly, we expect $M_x(t) \propto \sqrt{t\,\tau_a}$ growth to begin, exactly as in \Eq{eq:twogrowths}.
Although this behavior was derived using a simplified model, the full plane wave model obeys the same scalings, as can be confirmed numerically.

\section{Magnetic Flux from Axion Dark Matter}
\label{app:MFaDM}

In our results, we assumed a simple relation between the magnetic field ${\bf B}$ outside the sample that is measured at the pickup loop and the magnetization within the sample, in particular, $B_x = M_x$.
In general, an ${\cal O}(1)$ factor will correct this relation between the magnetization and the magnetic field.
In this appendix, we demonstrate this by explicitly computing that factor assuming a spherical sample.

The vector potential from such a setup is the sum of the contributions from a bulk and surface magnetization:
\begin{equation}
{\bf A}(t,{\bf r}) = \frac{1}{4\pi}\int d^3{\bf r}' \frac{[\nabla' \times {\bf M}({\bf r}',t) ]_{\rm ret} }{|{\bf r}-{\bf r}'|}
+ \frac{1}{4\pi} \oint_S dS^\prime \frac{[{\bf M}({\bf r}',t)]_{\rm ret} \times \hat{\bf n}'}{|{\bf r}-{\bf r}'|},
\label{eq:sampleA}
\end{equation}
where $\hat{\bf n}'$ is the unit vector perpendicular to the surface, and we evaluate the magnetization at the retarded time $t_{\rm ret} = t- \lvert{\bf r}-{\bf r}^\prime\rvert$.
Since the coherence length of axion dark matter for the range of masses we consider is much larger than the characteristic length scale of the experiment, the magnetization the axion induces in the sample will be spatially uniform.
In this case, the first term in \Eq{eq:sampleA} is parametrically suppressed, and the magnetic field induced by the sample will be sourced primarily by the surface term.

If the pickup loop is placed near the surface of the sample then $m_a|{\bf r} - {\bf r}'| \ll 1$ for all points ${\bf r}'$ such that,
\begin{equation}
{\bf A}(t,{\bf r}) \simeq \frac{1}{4\pi} \oint_S dS' \frac{{\bf M}(t)\times \hat{\bf n}'}{|{\bf r}-{\bf r}'|}.
\label{eq:Approx1}
\end{equation}
For a sample composed of a sphere of radius $R$, the vector potential and magnetic field outside the sample are those of a magnetic dipole,
\begin{equation}\begin{aligned}
{\bf A}(t,{\bf r}) &= \frac{R^3}{3} \frac{{\bf M}(t) \times \hat{\bf r}}{|{\bf r}|^3},\\
{\bf B}(t,{\bf r}) &= \frac{R^3}{3|{\bf r}|^3} \left[3({\bf M}(t)\cdot\hat{\bf r})\hat{\bf r}-{\bf M}(t)\right]\!.
\label{eq:dipoles}
\end{aligned}\end{equation}
In particular, if the magnetic field is measured close to the sample, and aligned such that we measure the field at ${\bf r} = R\, \hat{\bf x}$, then the measured magnetic field will be proportional to the magnetization: $B_x(t) = (2/3)M_x(t)$.
The relationship between the two is an ${\cal O}(1)$ value as claimed.

\section{Integration Time and Scan Strategy}
\label{app:inttime}

As shown in the main text, a resonant response is one of the characteristic features of the interaction of the axion with a sample of nuclear spins.
Given the narrow bandwidth of the response, in order to cover a range of possible axion masses, one must adjust the Larmor frequency by varying the external magnetic field.
One then needs to specify a scanning strategy for how long to integrate at each frequency and how far apart those frequencies should be placed.
Here we will provide further details of the scan strategy adopted in the main text and, in particular, detail the calculation of the total integration time.
In this discussion, we will continue to ignore the impact of the longitudinal relaxation time, $T_1$.
However, after $T_1$ the spins will lose their coherence in the direction proportional to the external magnetic field, and the sample will need to be repolarized.
We defer to \Refc{Budker:2013hfa} for further discussion.

By default, the scan strategy we adopt is to run until the sensitivity scaling slows to its lowest parametric growth.
From \Eq{eq:paramgrowth}, this occurs when $T = {\rm max}[\tau_a,T_2]$.
The second ingredient in the search strategy is how to spread the $\omega_0$ values scanned.
At high masses, when $\tau_a > T_2$, the width of the axion is larger than the resonant response of the sample.
Therefore, to ensure no axion is missed, we need to shift by a mass-dependent quantity, $\Delta \omega = 2\pi/\tau_a$.
At lower masses, the response of the instrument is broader than the axion, and therefore we should shift by this larger value $\Delta \omega = 2\pi/T_2$.
In summary, our integration time and $\Delta \omega$ both change on either side of $\tau_a = T_2$.

We can now determine the total integration time required in the high and low mass regimes.
Consider high masses first, where $\tau_a < T_2$.
Here we integrate by a constant time, but we have a variable $\Delta \omega$.
To describe this, it is convenient to rewrite the coherence time in terms of the equivalent quality factor for the axion, $\tau_a = 2 \pi/m_a v^2 = 2 \pi Q_a/m_a$.
For dark matter, $Q_a \sim v^{-2} \sim 10^6$, however, this allows us to generalize the discussion to more general driving sources (see, e.g., \Refc{Dror:2021nyr}).
If we scan a frequency range $[\omega_{\rm min},\omega_{\rm max}]$, then the total number of bins required is
\begin{equation}
N_{\rm high} = \frac{\ln(\omega_{\rm max}/\omega_{\rm min})}{\ln(1+Q_a^{-1})},	
\end{equation}
implying the high-mass integration time is,
\begin{equation}
T_{\rm high} = T_2 	\frac{\ln(\omega_{\rm max}/\omega_{\rm min})}{\ln(1+Q_a^{-1})}.
\end{equation}

At low masses, where $\tau_a > T_2$, we have a constant $\Delta \omega$, but a variable $T$.
Accordingly, the number of bins to cover a range $[\omega_{\rm min},\omega_{\rm max}]$ can be immediately written as,
\begin{equation}
N_{\rm low} = \frac{T_2}{2\pi} (\omega_{\rm max} - \omega_{\rm min}).
\end{equation}
The total integration time can then be written in terms of the digamma function, $\psi(x)$, as
\begin{equation}
T_{\rm low} = \frac{\omega_{\rm max} T_2^2}{2\pi} \left[ \psi \left( \frac{T_2 \omega_{\rm max}}{2\pi} \right) - \psi \left( \frac{T_2 \omega_{\rm min}}{2\pi} \right) \right]\!.	
\end{equation}

These expressions can now be used to evaluate the total integration time.
For the magnetic dipole operator, we considered two results in the main body, one for Xenon and the other for Helium.
In both cases, we took $T_2= 100~{\rm sec}$, implying that in both cases, the transition between the low and high mass regime occurs at $m \simeq 41.4~{\rm peV}$.
At the high end, the maximum frequency varies given the different assumed $B_{\rm max}$ in addition to the larger Helium-3 magnetic moment.
Combining these details, we confirm the integration times quoted in the main body,
\begin{equation}
{\rm magnetic~dipole\!:}\hspace{0.5cm}T_{\rm low} + T_{\rm high}
\simeq \left\{ \begin{array}{lc}
26.4 + 29.7~{\rm years} \simeq 56.1~{\rm years} & ({\rm Xenon}), \\[0.1cm]
26.4 + 35.1~{\rm years} \simeq 61.5~{\rm years} & ({\rm Helium}).
\end{array}\right.
\label{eq:Ttotal-MD}
\end{equation}
These times are significant, although they follow from the strategy suggested in \Refc{Graham:2013gfa}.
Given that we find the sensitivity is enhanced at high masses and reduced at lower values, if we instead only run from the maximum mass down to $1~{\rm neV}$ ($10~{\rm neV}$), then the total integration time would reduce to 19.6 years (12.3 years) for Xenon, and 25.0 years (17.7 years) for Helium.
We note that the ``QCD axion targeted'' scan we show in \Fig{fig:sensitivity} only used a decade of integration time.

We can further determine the total integration time required for the electric dipole operator, considered in App.~\ref{app:EDM}.
As discussed there, we consider the two phases suggested in \Refc{Budker:2013hfa}.
The first phase has $T_2 = 1~{\rm msec}$, whereas the second has $T_2 = 1~{\rm sec}$.
The corresponding masses where $\tau_a = T_2$ are $4.14~\mu{\rm eV}$ and $4.14~{\rm neV}$, respectively.
This transition regime for phase one is at such a large mass that, in fact, the entire integration occurs in the low-mass regime.
The integration times are now,
\begin{equation}
{\rm electric~dipole\!:}\hspace{0.5cm}T_{\rm low} + T_{\rm high}
\simeq \left\{ \begin{array}{lc}
5.41 + 0~{\rm hours} \simeq 5.41~{\rm hours} & ({\rm Phase}~1), \\[0.1cm]
152 + 60.3~{\rm days} \simeq 213~{\rm days} & ({\rm Phase}~2).
\end{array}\right.
\end{equation}
The integration times here are significantly shorted than those in \Eq{eq:Ttotal-MD}.
We note, however, that this is assuming integration times of $T = {\rm max}[\tau_a,T_2]$.
In \Refc{Budker:2013hfa}, they claim to use $T = 100~{\rm sec} \times {\rm max}[1,\tau_a/2T_2]$, and if this was adopted, we find that Phases 1 and 2 would run for roughly 120 and 40 years, respectively.
Nevertheless, we suspect this is not what was adopted in their results, as with this scan strategy our results in \Fig{fig:EDM} would look significantly different and we would no longer find close agreement for Phase 1.

\end{document}